\documentclass[%
 aip,
 sd,%
 amsmath,amssymb,
 reprint,%
]{revtex4-1}

\usepackage{graphicx}
\usepackage{dcolumn}
\usepackage{bm}

\begin{document}

\preprint{AIP/123-QED}

\title{Quantum coherence between cavity and artificial atom in a superconducting circuit QED ladder system}
\author{Qichun Liu}
\author{Han Cai}
\author{Yingshan Zhang}
\author{Jianshe Liu}
\author{Wei Chen}
\homepage{weichen@tsinghua.edu.cn}
\affiliation{%
Department of Microelectronics and Nanoelectronics, Institute of Microelectronics, Tsinghua National Laboratory of Information Science and Technology, Tsinghua University, Beijing 100084, China
}%

\date{\today}

\begin{abstract}
We have created a quantum three-level ${\rm \Xi}$ system with the cavity dispersive energy level in a superconducting circuit quantum electrodynamics system consisting of a transmon qubit and a cavity, and have directly observed the Autler-Townes splitting effect instead of representing it by the probability of the qubit being at each level. A coupler tone is applied on the transition between the second excited state of transmon and cavity dispersive level, while the cavity spectrum is probed. A doublet transmission and anormalous dispersion spectrum of the cavity level is clearly shown. The inverse Fourier transform of cavity spectrum indicates that there is a quantum coherence Rabi oscillation of the populations between cavity and qubit.
\end{abstract}

\maketitle

\section{Introduction}
\label{intro}

In a superconducting circuit quantum electrodynamics (QED) system\cite{QED Hua M}, one usually takes quantum levels of the qubit as useful levels and uses the cavity only as a control or measurement tool. In previous report, when the system is used to demonstrate the Autler-Townes splitting (ATS)\cite{ATS Autler} quantum optics effect, which happens when a quantum three-level system is driven with a strong resonant coupler tone, both of the coupler and probe tone interact with qubit levels\cite{ATS Cho SU,ATS Novikov S,ATS Suri B,ATS Li J,ATS Kelly WR,ATS Sillanpaa MA,ATS Baur M}. In order to characterize the ATS phenomenon, another measurement tone has to be applied to the system for measuring the qubit state and the ATS effect is represented by the probability of the qubit being at each level.

However, the cavity dispersive level also can be used as a quantum level. In a recent work\cite{realization EIT Novikov S}, S. Novikov et al create a three-level ${\rm \Lambda}$ system with cavity dispersive level. Here, we have constructed a ${\rm \Xi}$ quantum system with the cavity dispersive level and shown the ATS effect of cavity spectrum. As shown in Fig.\,1(a), a coupler tone is applied between qubit second excited state and the cavity dispersive level, while the cavity dispersive level is probed by a probe tone directly. This scheme gives an ATS phenomenon directly represented by probe tone, instead of by the qubit state. That means the output of probe tone is directly measurable and usable, by using the ATS effect in Fig.\,1(a) to construct a microwave router\cite{router Hoi I-C}.

In this paper, we focus on the spectrum of cavity level instead of the qubit quantum levels and report an experiment on direct observation of the ATS effect in a superconducting circuit QED ${\rm \Xi}$ system. We research into the inverse Fourier transform of the spectrum and show the quantum coherence Rabi oscillation of the populations between cavity and qubit. Our experimental data fits well with the theoretical calculation result of the driven cQED system model.

\section{experiment}
\label{experiment}

The system is consisted of a three-dimensional (3D) microwave cavity and a tunable SQUID type transmon\cite{3D transmon Rigetti C,3D transmon Paik H}. The transmon is fabricated on silicon dioxide substrate and the Al/AlOx/Al junction each has an area of 100 nm$\times$250 nm. The SQUID loop is of size 2.5 $\mu$m$\times$3.5 $\mu$m and each transmon shunting capacitor pad has an area of 250 $\mu$m$\times$500 $\mu$m. The transmon is embedded in the center of 3D copper cavity and the effective Josephson energy can be tuned by a global magnetic bias. The experiment is performed at 20 mK in a dilution refrigerator. For this experiment, the effective Josephson energy $E_{J}$ is tuned to be $E_{J}/h=42.418$ GHz and the charging energy $E_{C}$ is measured to be $E_{C}/h=259$ MHz. The bare cavity frequency is $\omega_{\rm cav}=2\pi\times8.121$ GHz and the coupling strength between the cavity and the transmon is $g=2\pi\times182$ MHz. We use $|ij\rangle$ ($i=g,e,f$ for trasmon states and $j=0,1$ for cavity states) to denote the eigenstates of the coupled system [see Fig.\,1(a)]. The dispersive frequency of cavity $\omega_{\rm g1,g0}=2\pi\times8.0870$ GHz and the cavity decay rate $\kappa$ is measured to be $\kappa=2\pi\times1.26$ MHz. The transition frequency between $|e0\rangle$ and $|g0\rangle$ is $\omega_{\rm e0,g0}=2\pi\times9.1160$ GHz and two photon transition frequency between $|f0\rangle$ and $|g0\rangle$ is $\omega_{\rm f0,g0}/2=2\pi\times8.9865$ GHz and the damping rate $\gamma=2\pi\times1.18$ MHz obtained by fitting the spectrum with a Lorentzian.

\begin{figure}
\centerline{\scalebox{0.6}{\includegraphics{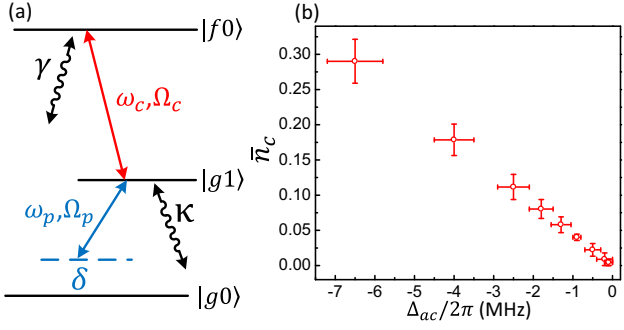}}}
\caption{\label{fig:epsfig1} (a)The model of energy levels comprising the Ladder system. It is driven by a coupler tone (red) and a probe tone (blue), with frequency being $\omega_{c}$ and $\omega_{p}$, and corresponding strength being $\Omega_{c}$ and $\Omega_{p}$. $\gamma$ is the damping rate of qubit level $|f0\rangle$ and $\kappa$ is the decay rate of cavity dispersive level $|g1\rangle$. (b)The off-resonant ac-Stark shift $\Delta_{ac}$ of $|f0\rangle$ caused by coupler tone. It is linear dependent on photons \={n}$_{c}$.}
\end{figure}

In experiment, the Ladder system is formed with states $|g0\rangle$, $|g1\rangle$ and $|f0\rangle$. A coupler tone is resonant with the transition between $|f0\rangle$ and $|g1\rangle$ . ($\omega_{c}=\tilde{\omega}_{\rm f0,g1}=\omega_{\rm f0,g1}+\Delta_{\rm ac}$. The $\Delta_{\rm ac}$ term represents the off-resonant ac-Stark shift of transmon state $|f0\rangle$ resulting from the coupler tone.) A probe tone is slightly detuned from the dispersive level of cavity ($\omega_{p}=\omega_{\rm g1,g0}+\delta$). We populate the cavity with different number of photons by probe tone, and fit a Poisson distribution of Lorentzian to the photon number splitting spectrum of the transmon\cite{number splitting Schuster DI}. The dispersive shift resulting from cavity-qubit coupling is calibrated to be $\chi_{\rm shift}/2\pi=-11.2$ MHz and the average photon number of probe tone is \={n}=0.16 for the work presented in this paper, corresponding to the probe tone strength $\Omega_{p}/2\pi=0.252$ MHz. The average photon number of coupler tone \={n}$_{c}$ is calibrated by the linear dependence of $\Delta_{\rm ac}$ on \={n}$_{c}$ ($\Delta_{\rm ac}=2\chi_{\rm shift}$\={n}$_{c}$)\cite{stark shift Gambetta J}. As shown in Fig.1\,(b), \={n}$_{c}$ is from 0.0044 to 0.286 for $\Delta_{\rm ac}/2\pi$ from $-0.1$ MHz to $-6.5$ MHz.

For Poisson distribution, when average photon number \={n}=0.16, we have P$(n=2)$=0.011, which is much smaller than the experimental noise $\pm 4$$\%$. Taking the approximation that the probe tone only involves $|g0\rangle$ and $|g1\rangle$, the driven system Hamiltonian becomes\cite{ATS Cho SU,realization EIT Novikov S}

\begin{eqnarray}
H&=&\nonumber\hbar[-\delta(|g1\rangle\langle g1|+|f0\rangle\langle f0|)+\frac{\Omega_{p}}{2}(|g1\rangle\langle g0|
\\&+&|g0\rangle\langle g1|)+\frac{\Omega_{c}}{2}(|f0\rangle\langle g1|+|g1\rangle\langle f0|)]\mathrm{.}
\end{eqnarray}

The quantum dynamics of the system can be described by the Lindblad master equation with density matrix $\rho$:
\begin{equation}
\frac{d\rho}{dt}=\frac{i}{\hbar}[H,\rho]+\sum_{j}(\mathcal{L}_{j}\rho\mathcal{L}_{j}^{\dag}-\frac{1}{2}\{\rho,\mathcal{L}_{j}^{\dag}\mathcal{L}_{j}\})\mathrm{.}
\end{equation}

The jump operators $\mathcal{L}_{j}$ contains the decay of state $|g1\rangle$ and the damping of state $|f0\rangle$ with $\kappa=2\pi\times1.26$ MHz and $\gamma=2\pi\times1.18$ MHz. Starting from the ground state $|g0\rangle$, Eq.(2) is solved numerically in steady state.

\section{results and discussion}

\begin{figure}
\centerline{\scalebox{0.316}{\includegraphics{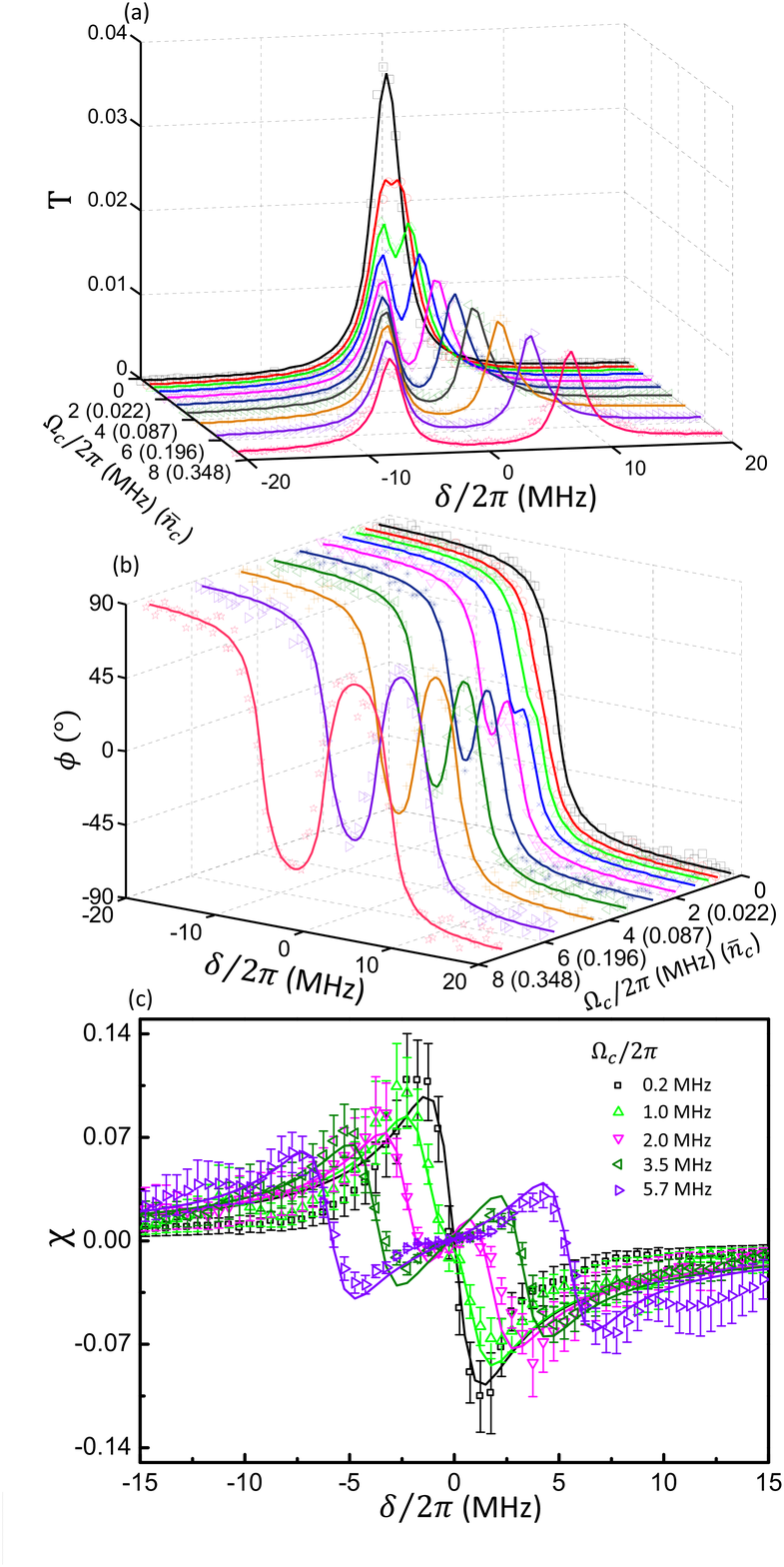}}}
\caption{\label{fig:epsfig2} The spectrum of cavity dispersive level for coupler tone strength $\Omega_{c}/2\pi$ ranging from $0.2$ MHz to $7.3$ MHz (corresponding to photon numbers \={n}$_{c}$ ranging from 0.0044 to 0.286). The experimental results are represented by symbols while the solid curves are numerically calculated with Eq.(2). For a clear vision, only five curves of the result are shown in dispersion spectrum $\chi$.}
\end{figure}

To characterize the ATS effect of the cavity dispersive level, we have measured the cavity spectrum at different coupler tone strength. Fig.\,2(a) shows the normalized cavity transmission amplitude T (indicated by symbols) at different coupler tone photon number \={n}$_{c}$, ranging from 0.0044 to 0.286, corresponding the coupling strength $\Omega_{c}/2\pi$ ranging from $0.2$ MHz to $7.3$ MHz. With coupling strength increasing, a doublet structure of T is clearly shown, and the doublet space is twice the coupling strength $\Omega_{c}$. The cavity transmission amplitude T is proportional to the photons $\langle a^{\dag}a\rangle=|\langle g1|\rho|g1\rangle|$ (indicated by solid curves) leaking out off the cavity, which is calculated with Eq.(2). Fig.2\,(b) shows the cavity transmission phase under the same condition as Fig.2\,(a). The theoretical phase $\phi$ curve is calculated by
 $\tan(\phi)={\rm Re}(\rho_{\rm g1,g0})/{\rm Im}(\rho_{\rm g1,g0})$, with the off-diagonal density matrix element $\rho_{\rm g1,g0}=\langle g1|\rho|g0\rangle$. We can see that both the cavity transmission amplitude T and phase $\phi$ can be well described by theoretical calculated results from Eq.(2). The dispersion spectrum of cavity $\chi$ (shown in Fig.2\,(c)) is calculated with transmission amplitude T and phase $\phi$ ($\chi={\rm T}\times \tan(\phi)$). Here, $\pm 4\%$ of experimental result as the noise is taken into account, corresponding to $\pm0.002$ for T and $\pm 7^{\circ}$ for $\phi$. The solid curves in Fig.2\,(c) represent ${\rm Re}(\rho_{\rm g1,g0}$). Two anormalous dispersion spectrum windows appear as the coupling strength increase, indicating that the cavity dispersive level splits into two.

\begin{figure}
\centerline{\scalebox{0.69}{\includegraphics{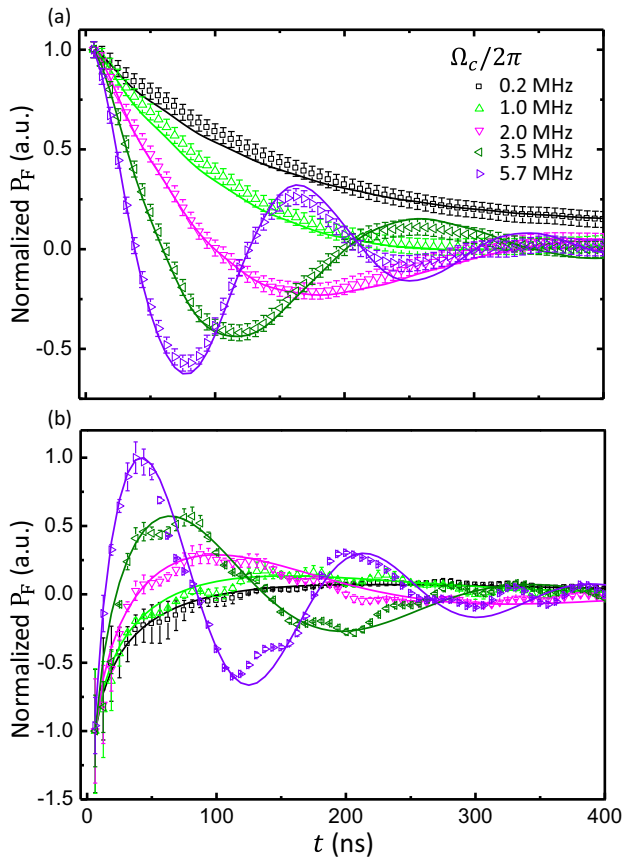}}}
\caption{\label{fig:epsfig3} The inverse Fourier transform of cavity transmission and dispersion spectrum ${\rm P}_{\rm F}$. (a)${\rm P}_{\rm F}=F^{-1}({\rm T})$, (b)${\rm P}_{\rm F}=F^{-1}(\chi)$. In ATS regime, ${\rm P}_{\rm F}\sim e^{-\kappa t}\cos(\Omega_{c}t+\theta)$ exhibits a quantum coherence Rabi oscillation between the states subject to the coupler tone.}
\end{figure}

The ATS effect is caused by electromagnetic pumping doublet structure effect, which means there is a quantum coherence Rabi oscillation of the populations between the states subject to the coupler tone\cite{quantum coherence Tohru Oishi,quantum coherence Lu XG}. This can be reflected in the inverse Fourier transform of transmission (or dispersion) spectrum ${\rm P}_{\rm F}=F^{-1}({\rm T})$ (or ${\rm P}_{\rm F}=F^{-1}(\chi)$). We show the experimental normalized results of ${\rm P}_{\rm F}=F^{-1}({\rm T})$ in Fig.3\,(a) and ${\rm P}_{\rm F}=F^{-1}(\chi)$ in Fig.3\,(b), represneted by symbols with errorbar resulting from the error of T and $\chi$. In Fig.3\,(a) and (b), the solid curves are theoretically calculated ${\rm P}_{\rm F}=F^{-1}(\langle a^{\dag}a\rangle)$ and ${\rm P}_{\rm F}=F^{-1}({\rm Re}(\rho_{\rm g1,g0}))$ with Eq.(2) respectively. While the cavity dispersive level splits into a doublet structure, a clear oscillation shows up in ${\rm P}_{F}$. The amplitude damping of the oscillation is because of the damping of state $|g1\rangle$, and the oscillation frequency (Rabi frequency) is equal to coupling strength $\Omega_{c}$. The oscillation can be described with ${\rm P}_{\rm F}\sim e^{-\kappa t}\cos(\Omega_{c}t+\theta)$.

\begin{figure}
\centerline{\scalebox{0.36}{\includegraphics{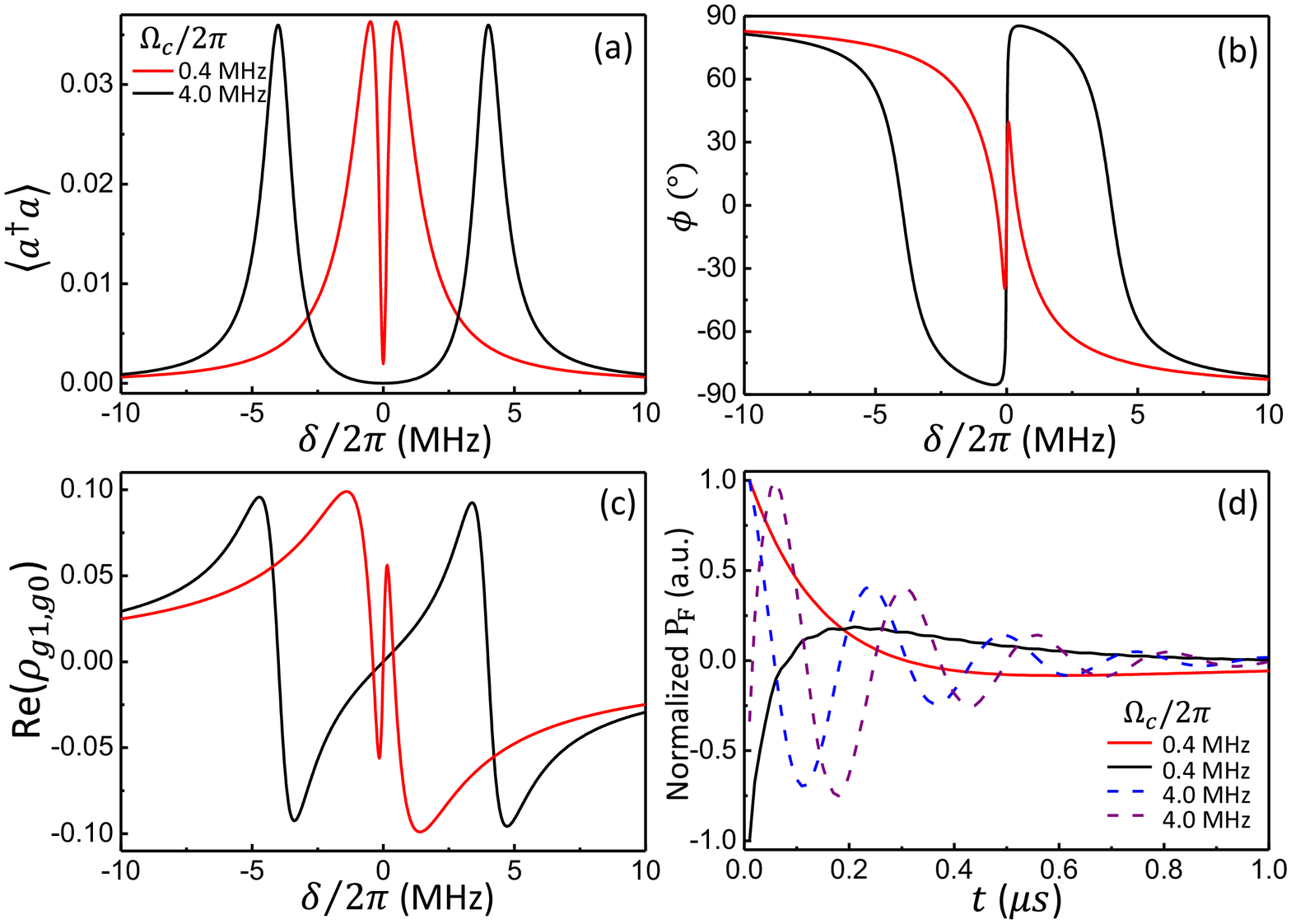}}}
\caption{\label{fig:epsfig4} Theoretical simulation of the quantum coherence effect between cavity and qubit with $\kappa/2\pi=1.26$ MHz and $\gamma/2\pi=0.02$ MHz in EIT regime ($\Omega_{c}/2\pi=0.4$ MHz) and in ATS regime ($\Omega_{c}/2\pi=4.0$ MHz).}
\end{figure}

The electromagnetically induced transparency (EIT) phenomenon\cite{EIT Marangos JP} is another important quantum optics effect in the driven three-level cQED system, and is closely related to ATS effect \cite{ATS-EIT Li HC,ATS-EIT Anisimov PM}. Because the EIT phenomenon is caused by destructive interference between two different excitation pathways, it can be used to slow down or even trap optical\cite{trap optical Hau LV} and microwave photons\cite{trap microwave Zhou X}, showing great potential application in single photon storage\cite{photon storage liu C}. Under different damping rate conditions, the driven system can transit between ATS regime and EIT regime. A significant EIT phenomenon can be realized by using the quantum coherence effect between cavity and qubit when the qubit's damping rate is much smaller than cavity's. We set the qubit's damping rate to be $\gamma=2\pi\times0.02$ MHz (corresponding to decay time $T_{1}=25$ $\mu$s, which can be reached in various superconducting qubit systems\cite{3D transmon Rigetti C,xmon Barends R,fluxnion Pop IM}), and all the other parameters of the system are the same as in this work. With this setup, the system can fulfill the condition for realizing EIT\cite{EIT condition Sun HC}. In EIT regime, the transparency window of spectrum is caused by destructive interference, and the quantum coherence Rabi oscillation of the populations between the states subjected to coupler tone is strongly suppressed by the state damping. That means the inverse Fourier transform of the spectrum ${\rm P}_{\rm F}$ is exponential decay curve, without oscillation.

The results of solving Eq.(2) with $\Omega_{c}/2\pi=0.4$ MHz (red curves in (a)-(c), solid curves in (d)) and $\Omega_{c}/2\pi=4.0$ MHz (black curves in (a)-(c), dash curves in (d)) respectively are shown in Fig.4\,. For $\Omega_{c}/2\pi=0.4$ MHz, an EIT window is significantly shown in Fig.4\,(a), and the inverse Fourier transform of the spectrum ${\rm P}_{\rm F}$ are simple exponential decay curves (solid curves in Fig.4\,(d)). For $\Omega_{c}/2\pi=4$ MHz, a doublet structure of the spectrum appears and ${\rm P}_{\rm F}$ becomes damping oscillations (dash curves in Fig.4\,(d)).

\section{conclusions}
\label{conclusions}

Taking the cavity dispersive state as a quantum level, we have create a quantum three-level ${\rm \Xi}$ system with it, and have directly observed the Autler-Townes splitting effect on the cavity spectrum instead of on the qubit quantum level. Both the transmission and dispersion spectrum of the cavity show a clear transition from one resonant window into two, and the inverse Fourier transform of them exhibit a significant quantum coherence Rabi oscillation of the populations between cavity and qubit. The experimental results can be well described by the master equation model of this driven superconducting circuit QED system. Furthermore, theoretical calculation indicates that a significant EIT phenomenon can be realized by this kind of quantum coherence effect between cavity and superconducting qubit when the qubit's damping rate is much smaller than cavity's.
\\
\\

This work is supported by the MOST 973 Program Grant Nos. 2011CBA00304, 2014CB848700 and 2014CB921401.
\\
\vspace*{0.8\baselineskip}
\hskip 12pt {\footnotesize%

{}

\end{document}